# Duffin-Kemmer-Petiau equation in curved space-time with scalar linear interactions

H. Hassanabadi[1], M. Hosseinpour [1] and M. de Montigny[2*]

[1]*Physics Department, Shahrood University, Shahrood, Iran P. O. Box: 3619995161-316, Shahrood, Iran*
[2]*Faculté Saint-Jean, University of Alberta, Edmonton, Alberta, T6C 4G9, Canada*

[*]*Corresponding author: mdemonti@ualberta.ca*

**Abstract**

In this paper, we study the covariant Duffin-Kemmer-Petiau (DKP) equation in the space-time generated by a cosmic string and we examine the linear interaction of a DKP field with gravitational fields produced by topological defects and thus study the influence of topology on this system. We highlight two classes of solutions defined by the product of the deficit angle with the angular velocity of the rotating frame. We solve the covariant form of DKP equation in an exact analytical manner for node-less and one-node states by means of an appropriate ansatz.

## 1. Introduction

The Duffin-Kemmer-Petiau (DKP) equation is a linear wave equation that enabled theoretical physicists to investigate both spin-zero and spin-one fields with a single equation in the relativistic regime [1–4]. It is a direct generalization of the Dirac equation based on the so-called DKP algebra [5]. This algebra admits three irreducible representations more common utilized in physics: a one-dimensional trivial representation, a five-dimensional representation for spin-zero fields, and a ten-dimensional representation that describes spin-one fields [6]. The DKP equation has been investigated in space-time with minimal length [7, 8], non-commutative phase space [9-12] and topological defects [13]. Such studies were motivated by the successful outcomes of DKP theory in various physical fields including particle and nuclear physics [1–5]. Recently, there has been an increased interest in the so-called DKP oscillator [14-23], in particular in the background of a magnetic cosmic string [13]. From the field-theoretical point of view, a cosmic string can be viewed as a consequence of symmetry breaking phase transition in the early universe [24]. Cosmic strings are among the most important class of linear topological defects with the conical geometry. In quantum field theory, the conical topology of the space-time due to the presence of a cosmic string causes a number of interesting physical effects. Until now, some problems have been investigated in curved space–time including the one-electron atom problem [25-27]. The dynamics of non-relativistic particles in curved space–time is also considered in Refs [28- 32].

In this work, we shall consider the quantum dynamics of scalar bosons with the DKP equation in the curved space-time of a cosmic sting. We will obtain the solutions of the equation in the presence of scalar linear interactions. We will analyze in detail the influence of the topological defect in the equation of motion, the DKP spinor and its energy spectrum. Note that the author Ref [33] also examined the non-inertial effects of rotating frames on the quantum dynamics of DKP scalar bosons embedded in a cosmic string background. But this was done (as well as in other references, such as [34] and others) by introducing a non-minimal substitution without any other interactions and obtaining in a usual manner the equations of motion, their solutions, energy, DKP spectrum, etc. In the present manuscript, we consider a similar problem with a scalar interaction, which we take as linear, that is $U = qr$. A linear potential, proportional to the radial coordinate $r$, has been used, for instance, to describe the confinement of quarks in a hadron and to calculate the bound states of quarks. In this context, several numerical calculations have been performed with regard to the nonrelativistic Schrödinger equation with a linear potential added to a Coulomb-type potential [35-38]. The Dirac equation with a linear potential has been discussed in Ref. [39]. The three-dimensional Dirac equation [40] and Klein–Gordon equation [41] with time-independent linear potentials have been studied and the bound-state exact solutions were obtained.

The work is organized as follows. Sec. 2 describes the DKP equation in a cosmic string background in a rotating frame with a general potential. The relevant tetrad basis and spin connections are obtained. In Sec. 3, we introduction the DKP oscillator by means of a non-minimal substitution, for scalar bosons for a linear potential, and we obtain the radial equations that are solved in the next two sections for different conditions involving the deficit angle and the angular velocity of the rotating frame. We make concluding remarks in the last section.

## 2. The DKP equation in curved space-time

The cosmic string space-time is an object described by the line element [42]

$$ds^2 = -dT^2 + dR^2 + \alpha^2 R^2 d\phi^2 + dZ^2 \tag{1}$$

where $\alpha$ is the deficit angle that depends on the linear mass density of the cosmic string. The rotating frame is obtained by using the following coordinate transformation

$$T = t, R = r, \phi = \omega t + \varphi, Z = z \tag{2}$$

where $\omega$ is the constant angular velocity of the rotating frame. Thus the line element (1) becomes

$$ds^2 = -(1 - \omega^2 \alpha^2 r^2) dt^2 + 2\omega \alpha^2 r^2 d\varphi dt + dr^2 + \alpha^2 r^2 d\varphi^2 + dz^2, \tag{3}$$

which describes the background of a cosmic string in a rotating frame. It is defined in the interval $0 < r < r_o$, where $r_o = 1/\omega\alpha$, and for $r > r_o$, it corresponds to a particle located outside of the line cone. Hence the geometry above is equivalent to a hard-wall confining potential, thus imposing that the wavefunction vanish as $r$ approaches $r_0$. This also leads to two classes of solutions: (1) $\omega\alpha$ arbitrary but finite and (2) $\omega\alpha \ll 1$.

Hereafter, we consider the DKP equation for a scalar boson in a curved space-time for scalar interaction is given by [12]

$$(i\beta^{\mu}\nabla_{\mu} - (M + U))\psi = 0 \tag{4}$$

where $c = \hbar = 1$ and the covariant derivative is

$$\nabla_{\mu} = \partial_{\mu} - \Gamma_{\mu} \tag{5}$$

$M$ is the mass of the DKP field and $U$ is the potential, that we will choose as linear in Sec. 3. The affine connection is defined by [43, 44]

$$\Gamma_{\mu} = \frac{1}{2}\omega_{\mu ab}[\beta^{a}, \beta^{b}] \tag{6}$$

$$\beta^{\mu} = e^{\mu}_{a}\beta^{a} \tag{7}$$

where we choose the $5 \times 5$ beta-matrices as follows

$$\beta^0 = \begin{pmatrix} \theta & 0_{2\times3} \\ 0_{3\times2} & 0_{3\times3} \end{pmatrix}, \vec{\beta} = \begin{pmatrix} 0_{2\times2} & \vec{\rho} \\ -\vec{\rho}^T & 0_{3\times3} \end{pmatrix}, \text{ with} \tag{8}$$

$$\theta = \begin{pmatrix} 0 & 1 \\ 1 & 0 \end{pmatrix}, \rho^1 = \begin{pmatrix} -1 & 0 & 0 \\ 0 & 0 & 0 \end{pmatrix}, \rho^2 = \begin{pmatrix} 0 & -1 & 0 \\ 0 & 0 & 0 \end{pmatrix}, \rho^3 = \begin{pmatrix} 0 & 0 & -1 \\ 0 & 0 & 0 \end{pmatrix}.$$

In Eq. (7), $e^{\mu}_{a}$ denotes the tetrad basis that we choose as [34]

$$e^{\mu}_{a} = \begin{pmatrix} 1/\sqrt{1-\rho^2} & 0 & \omega\alpha r/\sqrt{1-\rho^2} & 0 \\ 0 & 1 & 0 & 0 \\ 0 & 0 & \sqrt{1-\rho^2}/\alpha r & 0 \\ 0 & 0 & 0 & 1 \end{pmatrix}, \rho = \omega\alpha r \tag{9}$$

For the specific tetrad basis (9), we find from Eq. (7) that the curved-space beta-matrices read

$$\beta^{\circ} = \frac{1}{\sqrt{1-\rho^2}}(\beta^{\bar{\circ}} + \omega\alpha r\beta^{\bar{2}}) = \frac{1}{\sqrt{1-\rho^2}}\left[\begin{pmatrix} 0 & 1 & 0 & 0 & 0 \\ 1 & 0 & 0 & 0 & 0 \\ 0 & 0 & 0 & 0 & 0 \\ 0 & 0 & 0 & 0 & 0 \\ 0 & 0 & 0 & 0 & 0 \end{pmatrix} + \omega\alpha r\begin{pmatrix} 0 & 0 & 0 & -1 & 0 \\ 0 & 0 & 0 & 0 & 0 \\ 0 & 0 & 0 & 0 & 0 \\ 1 & 0 & 0 & 0 & 0 \\ 0 & 0 & 0 & 0 & 0 \end{pmatrix}\right] = \frac{1}{\sqrt{1-\rho^2}}\begin{pmatrix} 0 & 1 & 0 & -\omega\alpha r & 0 \\ 1 & 0 & 0 & 0 & 0 \\ 0 & 0 & 0 & 0 & 0 \\ \omega\alpha r & 0 & 0 & 0 & 0 \\ 0 & 0 & 0 & 0 & 0 \end{pmatrix} \tag{10.a}$$

$$\beta^r = \beta^{\bar{1}} = \begin{pmatrix} 0 & 0 & -1 & 0 & 0 \\ 0 & 0 & 0 & 0 & 0 \\ 1 & 0 & 0 & 0 & 0 \\ 0 & 0 & 0 & 0 & 0 \\ 0 & 0 & 0 & 0 & 0 \end{pmatrix} \tag{10.b}$$

$$\beta^{\varphi} = \frac{\sqrt{1-\rho^2}}{\alpha r} \beta^{\bar{2}} = \frac{\sqrt{1-\rho^2}}{\alpha r} \begin{pmatrix} 0 & 0 & 0 & -1 & 0 \\ 0 & 0 & 0 & 0 & 0 \\ 0 & 0 & 0 & 0 & 0 \\ 1 & 0 & 0 & 0 & 0 \\ 0 & 0 & 0 & 0 & 0 \end{pmatrix},$$ (10.c)

$$\beta^{z} = \beta^{\bar{3}} = \begin{pmatrix} 0 & 0 & 0 & 0 & -1 \\ 0 & 0 & 0 & 0 & 0 \\ 0 & 0 & 0 & 0 & 0 \\ 0 & 0 & 0 & 0 & 0 \\ 1 & 0 & 0 & 0 & 0 \end{pmatrix}$$ (10.d)

The spin connections are given by

$$\Gamma_{\circ} = \frac{\omega\alpha}{\sqrt{1-\rho^2}} (\omega\alpha r[\beta^{\bar{0}},\beta^{\bar{1}}] - [\beta^{\bar{1}},\beta^{\bar{2}}]) = \frac{\omega\alpha}{\sqrt{1-\rho^2}} \begin{pmatrix} 0 & 0 & 0 & 0 & 0 \\ 0 & 0 & -\omega\alpha r & 0 & 0 \\ 0 & -\omega\alpha r & 0 & 1 & 0 \\ 0 & 0 & -1 & 0 & 0 \\ 0 & 0 & 0 & 0 & 0 \end{pmatrix}$$ (11.a)

$$\Gamma_{r} = \frac{-\omega\alpha}{1-\rho^2} [\beta^{\bar{0}},\beta^{\bar{2}}] = \frac{-\omega\alpha}{1-\rho^2} \begin{pmatrix} 0 & 0 & 0 & 0 & 0 \\ 0 & 0 & 0 & -1 & 0 \\ 0 & 0 & 0 & 0 & 0 \\ 0 & -1 & 0 & 0 & 0 \\ 0 & 0 & 0 & 0 & 0 \end{pmatrix},$$ (11.b)

$$\Gamma_{\varphi} = \frac{\Gamma_{\circ}}{\omega} = \frac{\alpha}{\sqrt{1-\rho^2}} \begin{pmatrix} 0 & 0 & 0 & 0 & 0 \\ 0 & 0 & -\omega\alpha r & 0 & 0 \\ 0 & -\omega\alpha r & 0 & 1 & 0 \\ 0 & 0 & -1 & 0 & 0 \\ 0 & 0 & 0 & 0 & 0 \end{pmatrix}$$ (11.c)

### 3. The DKP oscillator in a cosmic string background with a linear potential

The DKP oscillator is introduced via the non-minimal substitution [1,45]

$$\frac{1}{i}\vec{\nabla}_{\alpha} \rightarrow \frac{1}{i}\vec{\nabla}_{\alpha} - iM\varpi\eta_{\circ}\vec{r}$$

where $\varpi$ is the oscillator frequency, $M$ is the mass of the boson, and $\vec{\nabla}$ is defined in Eq. (5). Although we shall set $\varpi = 0$ in the next section, we keep this oscillator term in this section in order to compare our results with Refs. [13] and [33]. We consider only the radial component in the non-minimal substitution. As the interaction is time-independent, one can write $\Psi(r,t) = \Phi(r)\exp(-iEt)$, where $E$ is the energy of the scalar boson, in such a way that the time-independent DKP equation becomes

$$[\frac{1}{\sqrt{1-\rho^2}}\begin{pmatrix} 0 & E & 0 & -\omega\alpha rE & 0 \\ E & 0 & 0 & 0 & 0 \\ 0 & 0 & 0 & 0 & 0 \\ E\omega\alpha r & 0 & 0 & 0 & 0 \\ 0 & 0 & 0 & 0 & 0 \end{pmatrix} + \begin{pmatrix} 0 & 0 & -i(\partial_r - M\varpi r) & 0 & 0 \\ 0 & 0 & 0 & 0 & 0 \\ i(\partial_r + M\varpi r) & 0 & 0 & 0 & 0 \\ 0 & 0 & 0 & 0 & 0 \\ 0 & 0 & 0 & 0 & 0 \end{pmatrix} + i\frac{\sqrt{1-\rho^2}}{\alpha r}\begin{pmatrix} 0 & 0 & 0 & -\partial_\varphi & 0 \\ 0 & 0 & 0 & 0 & 0 \\ 0 & 0 & 0 & 0 & 0 \\ \partial_\varphi & 0 & 0 & 0 & 0 \\ 0 & 0 & 0 & 0 & 0 \end{pmatrix}$$ (12)

$$-i\frac{1}{r}\begin{pmatrix} 0 & 0 & 1 & 0 & 0 \\ 0 & 0 & 0 & 0 & 0 \\ 0 & 0 & 0 & 0 & 0 \\ 0 & 0 & 0 & 0 & 0 \\ 0 & 0 & 0 & 0 & 0 \end{pmatrix} + i\begin{pmatrix} 0 & 0 & 0 & 0 & -\partial_z \\ 0 & 0 & 0 & 0 & 0 \\ 0 & 0 & 0 & 0 & 0 \\ 0 & 0 & 0 & 0 & 0 \\ \partial_z & 0 & 0 & 0 & 0 \end{pmatrix} - (M+U)I_{5\times 5}]\Phi = 0$$

The five-component DKP spinor can be written as $\Phi^T = (\Phi_1, \Phi_2, \Phi_3, \Phi_4, \Phi_5)$ and the DKP equation (4) becomes

$$-(M+U)\Phi_1 + \frac{1}{\sqrt{1-\rho^2}}E\Phi_2 - i\left(\partial_r - M\varpi r + \frac{1}{r}\right)\Phi_3 - \left(\frac{1}{\sqrt{1-\rho^2}}\omega\alpha rE + i\frac{\sqrt{1-\rho^2}}{\alpha r}\partial_\varphi\right)\Phi_4 - i\partial_z\Phi_5 = 0 \quad (13.1)$$

$$\frac{1}{\sqrt{1-\rho^2}}E\Phi_1 - (M+U)\Phi_2 = 0 \tag{13.2}$$

$$i(\partial_r + M\varpi r)\Phi_1 - (M+U)\Phi_3 = 0 \tag{13.3}$$

$$\left(\frac{1}{\sqrt{1-\rho^2}}E\omega\alpha r + i\frac{\sqrt{1-\rho^2}}{\alpha r}\partial_\varphi\right)\Phi_1 - (M+U)\Phi_4 = 0 \tag{13.4}$$

$$i\partial_z\Phi_1 - (M+U)\Phi_5 = 0 \tag{13.5}$$

By considering $\Phi_1 \propto e^{im\varphi}e^{ikz}F(r)$, we obtain the following equation of motion for the first component $\Phi_1$ of the DKP spinor:

$$[\partial_r^2 + \left(\frac{-\partial_r U}{(M+U)} + \frac{1}{r}\right)\partial_r - $$
$$\frac{\partial_r U}{(M+U)}(M\varpi r) - \frac{1-(\omega\alpha r)^2}{\alpha^2 r^2}m^2 + 2\omega Em - k^2 - (M+U)^2 + E^2 + 2\varpi M - M^2\varpi^2 r^2]F(r) = 0 \tag{14}$$

where $m$ is the magnetic quantum number, $k$ is the wave number and $M$ is the mass of the DKP boson. Let us take $F(r)$ as

$$F(r) = (M+U)^{1/2}r^{-1/2}R_{n,l}(r) \tag{15}$$

As mentioned in the introduction, we now specialize the scalar potential as being linear in $r$:

$$U(r) = qr$$

Then Eq. (14) changes to

$$R_{n,\ell}''(r) + \left[\frac{1/4 - m_\alpha^2}{r^2} + \frac{q}{2Mr} - 2Mqr \right.$$
$$\left. -(q^2 + M^2\varpi^2)r^2 - \frac{3q^2}{4(M+qr)^2} + \frac{-q^2 + 2M^3\varpi}{2M(M+qr)} + m_\alpha^2(\omega\alpha)^2 + M\varpi + \kappa^2\right]R_{n,\ell}(r) = 0, \tag{16}$$

where $\kappa^2 = E^2 - k^2 - M^2 + 2Em\omega$, $m_\alpha = \frac{m}{\alpha}$. In order to solve Eq. (16), we assume [46-48]

$$R_{n,l}(r) = f_n(r)e^{g_l(r)} \tag{17}$$

where

$$f_n(r) = \begin{cases} 1 & n = 0 \\ \prod_{i=1}^{n}(r - \alpha_i^n) & n \geq 1 \end{cases} \quad (18)$$

and $g_l(r)$ is given by

$$g_\ell = b_1 r + b_2 r^2 + b_3 \ln(r) + b_4 \ln(M + qr) \quad (19)$$

For node-less states, with $n = 0$, we have $R_{0,l}(r) = e^{g_l(r)}$ such that

$$R''_{0,l}(r) + (-g''_l - g'^2_l) R_{0,l}(r) = 0 \quad (20)$$

Then Eq (20) can be written as

$$R''_{o,\ell}(r) + \left[ -b_1^2 - 2b_2 - 4b_2 b_3 - 4b_2 b_4 - \frac{b_3(b_3 - 1)}{r^2} \right.$$
$$\left. - \frac{2(Mb_1 b_3 + b_3 b_4 q)}{Mr} - 4b_1 b_2 r - 4b_2^2 r^2 - \frac{b_4(b_4 - 1)q^2}{(M + qr)^2} + \frac{2(2b_2 b_4 M^2 - b_1 b_4 Mq + b_3 b_4 q^2)}{M(M + qr)} \right] R_{o,\ell}(r) = 0 \quad (21)$$

When we compare Eq. (16) with Eq. (21), we obtain seven equations

$$m_\alpha^2(\omega\alpha)^2 + M\varpi + \kappa^2 + b_1^2 + 2b_2 + 4b_2 b_3 + 4b_2 b_4 = 0$$
$$b_3(b_3 - 1) + \frac{1/4 - m_\alpha^2}{r^2} = 0$$
$$\frac{2(Mb_1 b_3 + b_3 b_4 q)}{M} + \frac{q}{2M} = 0$$
$$4b_1 b_2 - 2Mq = 0 \quad (22)$$
$$4b_2^2 - (q^2 + M^2\varpi^2) = 0$$
$$b_4(b_4 - 1) - \frac{3}{4} = 0,$$
$$2(2b_2 b_4 M^2 - b_1 b_4 Mq + b_3 b_4 q^2) + q^2 - 2M^3\varpi = 0$$

By solving Eq. (22) we find

$$b_2 = -\frac{1}{2}\sqrt{q^2 + M^2\varpi^2} \to b_1 = -\frac{Mq}{\sqrt{q^2 + M^2\varpi^2}}$$
$$b_2 = \frac{1}{2}\sqrt{q^2 + M^2\varpi^2} \to b_1 = \frac{Mq}{\sqrt{q^2 + M^2\varpi^2}}$$
$$b_3 = \frac{-2m + \alpha}{2\alpha}, b_3 = \frac{2m + \alpha}{2\alpha}$$
$$b_4 = -\frac{1}{2}, b_4 = \frac{3}{2} \quad (23)$$

As mentioned in Sec. 2, the peculiar behavior of this background is that it is defined in the interval $0 < r < r_o$, where $r_o = 1/\omega\alpha$, so that the problem presents two different classes of solutions that are determined by the value of $\omega\alpha$, according to whether is it finite and arbitrary or much smaller than 1. In absence of scalar potential, our result is covered in Ref. [33] where Castro investigated the DKP oscillator in cosmic-string background and considered the combined effects of a rotating frames and cosmic-string on the equation of motion. In the absence of a scalar potential, Eq. (14) is the same as in [33], and the solutions reduce to confluent hypergeometric functions; the solution is obtained for two cases: arbitrary $\omega\alpha$, and $\omega\alpha \ll 1$.

In the following section, we shall consider the two above-mentioned classes of solutions, in which case we obtain the node-less state ($n = 0$) and one-node state ($n = 1$). In all cases, we also consider the restriction that the DKP spinor must vanish at $r_0$.

### 4. Solution of DKP equation for node-less and one-node states with arbitrary $\omega\alpha$

In this section, we first proceed to find the solution of the DKP equation without the oscillator; that is, with $\varpi = 0$ in Eq. (16)

$$R''_{n,\ell}(r) + \left[\frac{1/4 - m_\alpha^2}{r^2} + \frac{q}{2Mr} - 2Mqr - q^2 r^2 - \frac{3q^2}{4(M+qr)^2} - \frac{q^2}{2M(M+qr)} + m_\alpha^2(\omega\alpha)^2 + \kappa^2\right] R_{n,\ell}(r) = 0, \qquad (24)$$

For node-less states, with $n = 0$, we compare Eqs. (24) and (21) and thus find

$$m_\alpha^2(\omega\alpha)^2 + \kappa^2 + b_1^2 + 2b_2 + 4b_2 b_3 + 4b_2 b_4 = 0$$
$$b_3(b_3 - 1) + 1/4 - m_\alpha^2 = 0$$
$$4(Mb_1 b_3 + b_3 b_4 q) + q = 0$$
$$4b_1 b_2 - 2Mq = 0 \qquad (25)$$
$$4b_2^2 - q^2 = 0$$
$$b_4(b_4 - 1) - 3 = 0$$
$$4(2b_2 b_4 M^2 - b_1 b_4 Mq + b_3 b_4 q^2) + q^2 = 0$$

From these equations, we find the following parameters

$$b_2 = +\frac{q}{2} \to b_1 = +M,$$
$$b_2 = -\frac{q}{2} \to b_1 = -M$$
$$b_3 = \frac{\alpha \pm 2m}{2\alpha}, \qquad (26)$$
$$b_4 = -\frac{1}{2}, \frac{3}{2}.$$

In order to find the eigenfunctions, we observe that the solution for $0 < r < r_o$ can be written as

$$\Phi_{0,l} = e^{im\varphi} e^{ikz} e^{b_1 r + b_2 r^2} r^{b_3} (M + qr)^{b_4} \qquad (27)$$

As observed in Sec. 2, because of the restriction on the radial coordinate caused by non-inertial effects, a physical solution is possible only if the eigenfunction vanishes at $r = r_o = 1/\omega\alpha$

$$R_{0,l}(r_o) = R_{0,l}(1/\omega\alpha_o) = e^{g_l(r)} = 0$$

The restriction leads to

$$M + qr_o = 0 \qquad (28)$$

In order to solve Eq. (24) for one-node state, $n = 1$, we assume

$$R_{n=1,l}(r) = f_1(r)e^{g_1(r)}$$

In this case, we have, from Eq. (18),

$$f_1(r) = r - \alpha_1^1 \qquad (29)$$

From (24) we find for the one-node state that

$$R_{1,1}(r)'' - \left(\frac{2g_1'(r)}{r - \alpha_1^1} + g_1'^2(r)\right) + g_1''(r))R_{1,1}(r) = 0. \qquad (30)$$

By substituting $g(r)$ into Eq. (30) and comparing with Eq. (19), we obtain

$$4b_2^2 - q^2 = 0; \qquad (31.1)$$
$$b_3(1 - b_3) + (m_\alpha^2 - 1/4) = 0; \qquad (31.2)$$
$$3Mq + 4b_4 Mq - 4b_4^2 Mq + 3\alpha_1^1 q^2 + 4\alpha_1^1 b_4 q^2 - 4\alpha_1^1 b_4^2 q^2 = 0 \qquad (31.3)$$
$$(4b_1 b_2 - 4\alpha_1^1 b_2^2 - 2Mq + \alpha_1^1 q^2) = 0 \qquad (31.4)$$

$$4b_2 b_4 M - 2b_1 q + \alpha_1^1 b_1^2 q + 2\alpha_1^1 b_2 q - 2b_1 b_3 q + 4\alpha_1^1 b_2 b_3 q - 2b_1 b_4 q + 4\alpha_1^1 b_2 b_4 q + \alpha_1^1 \kappa^2 q + \alpha_1^1 m_\alpha^2 \alpha^2 \omega^2 q = 0 \qquad (31.5)$$
$$-4m_\alpha^2 M + M + 4b_3 M - 8\alpha_1^1 b_1 b_3 M + 4b_3^2 M - 2\alpha_1^1 q - 8\alpha_1^1 b_3 b_4 q = 0 \qquad (31.6)$$
$$(b_1^2 + 6b_2 - 4\alpha_1^1 b_1 b_2 + 4b_2 b_3 + 4b_2 b_4 + \kappa^2 + 2\alpha_1^1 Mq + m_\alpha^2 \alpha^2 \omega^2) = 0 \qquad (31.7)$$
$$16b_2 b_4 M^3 - 8b_1 b_4 M^2 q + 16\alpha_1^1 b_2 b_4 M^2 q - Mq^2 + 4b_4 Mq^2 - 8\alpha_1^1 b_1 b_4 Mq^2 + 8b_3 b_4 Mq^2 + 4b_4^2 Mq^2 + 2\alpha_1^1 q^3 + 8\alpha_1^1 b_3 b_4 q^3 = 0 \qquad (31.8)$$

From Eqs. (31.1-4), we find

$$\begin{cases} b_2 = \dfrac{+q}{2} \to b_1 = M \\ b_2 = \dfrac{-q}{2} \to b_1 = -M \end{cases}, \quad \begin{cases} b_3 = \dfrac{2m + \alpha}{2\alpha} \\ b_3 = \dfrac{-2m + \alpha}{2\alpha} \end{cases}, \quad \begin{cases} b_4 = \dfrac{3}{2} \\ b_4 = \dfrac{-1}{2} \end{cases} \qquad (32)$$

Then the solution for $0 < r < r_o$ can be written as

$$\Phi_{1,1} = (r - \alpha_1^1)e^{im\varphi}e^{ikz}e^{b_1 r + b_2 r^2} r^{b_3} (M + qr)^{b_4} \qquad (33)$$

The restriction that the DKP spinor must vanish at $r_0$ implies that either $M + qr_o = 0$ or $r_o - \alpha_1^1 = 0$; but the latter is not acceptable for one-node states since the eigenfunction has zeros at $r = 0$, $\alpha_1^1 < r_o$ (at the node) and $r_0$ (the hard wall), implying that one must have $\alpha_1^1 \neq r_0$.

The node-less states are defined only when $b_1 = -M, b_2 = \dfrac{-q}{2}; b_3 = \dfrac{2m + \alpha}{2\alpha}; b_4 = \dfrac{3}{2}$ or $b_1 = M, b_2 = \dfrac{q}{2}; b_3 = \dfrac{2m + \alpha}{2\alpha}; b_4 = \dfrac{3}{2}$. In the following, we depict the node-less eigenfunctions of the system in terms of $r$ for the two cases above respectively. The eigenfunction is plotted as a function of $r$ in Figs. 1 and 2. Both graphs are plotted for $m = k = 1$. In both figures, the restriction condition are the same and thus the solution is restricted to the interval $0 \leq r \leq 2$. In one-node state, $\alpha_1^1$ should be a positive parameter. In Eq. (31.6) only for $b_2 = \frac{q}{2}, b_1 = M, b_3 = \frac{2m+\alpha}{2\alpha}, b_4 = \frac{3}{2}$ we can have a positive parameter for $\alpha_1^1$ and therefore a physical state. The eigenfunction of one-node states are plotted in Fig. 3 as a function of $r$ with the solution restricted to the interval $0 \leq r \leq 1.5$.

## 5. Solution of DKP equation for node-less and one-node states in limit $\omega\alpha \ll 1, r_\circ \to \infty$

Hereafter, the main changes are the boundary condition and the restriction that the DKP spinor is zero at the wall $r_0$. When we consider $r_\circ \to \infty$, in limit where $\omega\alpha = 1$, Eq. (24) is rewritten as

$$R_{n,1}''(r) + \left[\frac{1/4 - m_\alpha^2}{r^2} + \frac{q}{2Mr} - 2Mqr - q^2r^2 - \frac{3q^2}{4(M+qr)^2} - \frac{q^2}{2M(M+qr)} + \kappa^2\right] R_{n,1}(r) = 0 \tag{34}$$

$$\kappa^2 = E^2 - K^2 - M^2 + 2Em\omega, m_\alpha = \frac{m}{\alpha}$$

For node-less states, with $n = 0$, from Eqs. (34) and (21), we obtain

$$\kappa^2 + b_1^2 + 2b_2 + 4b_2b_3 + 4b_2b_4 = 0$$
$$b_3(b_3 - 1) + 1/4 - m_\alpha^2 = 0$$
$$4(Mb_1b_3 + b_3b_4q) + q = 0$$
$$4b_1b_2 - 2Mq = 0 \tag{35}$$
$$4b_2^2 - q^2 = 0$$
$$b_4(b_4 - 1) - 3 = 0$$
$$4(2b_2b_4M^2 - b_1b_4Mq + b_3b_4q^2) + q^2 = 0$$

This equation leads to the same parameters as Eq. (32). Thus the solution for $0 < r < \infty$ for node-less states can be written as

$$\Phi_{0,1} = e^{im\varphi}e^{ikz}e^{b_1r+b_2r^2}r^{b_3}(M+qr)^{b_4} \tag{36}$$

The physical states are defined only when the parameters are given by $b_1 = -M, b_2 = \frac{-q}{2}; b_3 = \frac{2m+\alpha}{2\alpha}; b_4 = \frac{3}{2}$ or $b_1 = -M, b_2 = \frac{-q}{2}; b_3 = \frac{2m+\alpha}{2\alpha}; b_4 = \frac{-1}{2}$. For one-node states, our results are the same as Eq. (31) and (32) with take into account $\omega\alpha = 1$. Then the solution for $0 < r < \infty$ can be written as

$$\Phi_{1,1} = (r - \alpha_1^1)e^{im\varphi}e^{ikz}e^{b_1r+b_2r^2}r^{b_3}(M+qr)^{b_4} \tag{37}$$

For one-node states, we find that only for $b_2 = \frac{-q}{2}, b_1 = -M, b_3 = \frac{2m+\alpha}{2\alpha}, b_4 = \frac{3}{2}$ we can have a positive parameter for $\alpha_1^1$ and therefore a physical state.

In the following, we depict the eigenfunction of the system in terms of $r$ and obtain the energy of node-less state for node-less and one-node states in the limit $\omega\alpha = 1$ respectively. The negative energy solutions are associated to the noninertial effect of rotating frames, which in turn indicate a Sagnac-type effect. Both particle and antiparticle energy levels are included in the spectrum and that result of the noninertial effect is to break the symmetry of the energy spectrum about $E = 0$ [49, 50]. Table 1 contains some energies of the node-less states with various values of $\omega\alpha$ and with $b_1 = -M, b_2 = \frac{-q}{2}; b_3 = \frac{2m+\alpha}{2\alpha}; b_4 = \frac{3}{2}$. Our Table 2 contains similar data except for $b_1 = -M, b_2 = \frac{-q}{2}; b_3 = \frac{2m+\alpha}{2\alpha}; b_4 = \frac{-1}{2}$. In Figs. 4 and 5, the eigenfunction is plotted as a function of $r$ for both values of $b_4$. Both graphs are plotted for $M = m = k = 1$ for $\alpha$ varying between 0.6 and 1.0. From Fig 4, in which $b_4 = 3/2$, we observe that as $\alpha$ increases, the peak shifts towards lower values of $r$ and the maximum value of $R_0$ decreases. Note that the eigenfunctions intersect at $r = 1$ for all values of $\alpha$. In Fig 5 $b_4 = -1/2$, the situation is somewhat reversed: as $\alpha$ increases the peak also shifts towards lower values of $r$ but in this case the maximum value of $R_0$ increases. Here also the eigenfunctions intersect at $r = 1$ for all values of $\alpha$. Fig. 6 illustrates the profile of the positive solution of energy as a function of $\alpha$. In the interval $0 < \alpha \leq 1$, it is seen that the energy decreases for increasing $\alpha$. Fig. 7 illustrates the profile of the negative solution of energy as a function of $\alpha$. In the interval $0 < \alpha \leq 1$, the energy increases for increasing $\alpha$. As expected, the energy tends to their behavior in flat space time for different $\omega$ when $\alpha$ tends toward 1. Figs 6 and 7 are plotted for different values of $\omega$ where $\omega = 0$ represent the non-rotated frame. In both Figs. 6 and 7, the numerical value of energy decreases for increasing $\omega$. The eigenfunction of one-node states are plotted as a function of $r$ for different $\alpha$ in Fig 8. The positive and negative solutions of energy for one-node states are plotted respectively in Fig 9 and 10 as a function of $\alpha$.

**Conclusions**

We studied the covariant DKP equation in the curved space-time of a cosmic string in a rotating reference frame in presence of a linear scalar interaction. We calculated the scalar-boson solutions of the DKP equation with polar coordinates and obtain the equation of motion of scalar bosons. We solved the DKP analytically for node-less and one-node states by using an ansatz to find the first component of the DKP spinor. We analyzed the effects of the angular velocity of the rotating frame $\omega$ and the angular deficit $\alpha$ of the cosmic string background, and found that the solutions are determined on the combined effect of $\omega$ and $\alpha$. This background is defined in the interval $0 \leq r \leq r_0 = \frac{1}{\omega\alpha}$, whereas for $r > r_0$, it corresponds to a particle located outside of the line cone. Thus the line element in Eq. (3) turns our to be equivalent to a hard-wall confining potential, thereby imposing that the DKP spinor be equal to zero at $r \rightarrow r_0$. This also suggests that there are two classes of solutions, depending on the product $\omega\alpha$. First, when the product $\omega\alpha$ has an arbitrary and finite value, we obtain the condition that makes the eigenfunction vanishes at $r = r_0$. In second case, we consider the condition that the product of $\omega\alpha \ll 1$ which is equivalent to the hard–wall at $r_0 \rightarrow \infty$.

For both classes of solutions, we obtained the eigenfunction for node-less and one-node states. We have also plotted the eigenfunction behavior graphically for node-less and one-node states respectively with $r_0 = 2$ and $r_0 = 1.5$. Also for $\omega\alpha \ll 1$ the related eigenfunctions are plotted.

We computed the energies $E_{0,\ell}$ and $E_{1,1}$ of this system for specific values of the solution parameters and for different values of $\omega\alpha \ll 1$. We also plotted that positive and negative values of $E_{0,\ell}$ and $E_{1,\ell}$. Besides exhibiting the usual node structure, with the number of nodes equal to *n* + 1, the graphs of $E_{0,\ell}$ and $E_{1,\ell}$ are, as expected, non-symmetric because of the linear potential, in a manner reminiscent of the Airy functions encountered when we solve the Schrödinger equation in a similar potential. The Figs. 1 and 2 display some sort of left-right reversal because the two choices of the parameter $b_2$ amount to changing the sign of the potential's slope. Figs. 6 and 7 show that $|E_{0,\ell}|$ decreases as $\alpha$ increases. Similarly, Figs. 9 and 10 show that $|E_{1,\ell}|$ decreases as $\omega\alpha$ increases.

## Acknowledgment

It is a great pleasure for the authors to thank the referees and L.B. Castro for useful comments on the manuscript. M. de Montigny acknowledges financial support from the Natural Sciences and Engineering Research Council (NSERC) of Canada.

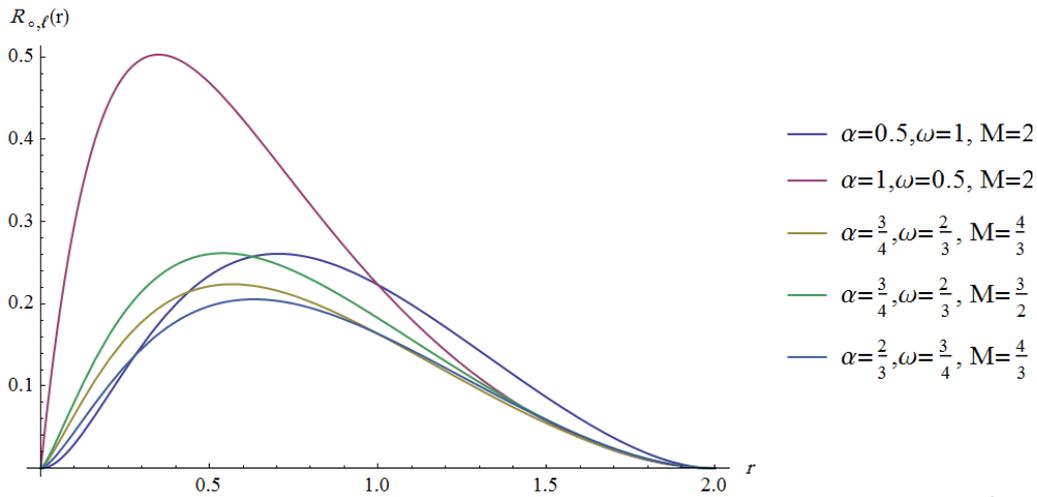

Fig. 1. The eigenfunctions for $b_1 = -M, b_2 = \frac{-q}{2}; b_3 = \frac{2m+\alpha}{2\alpha}; b_4 = \frac{3}{2}$

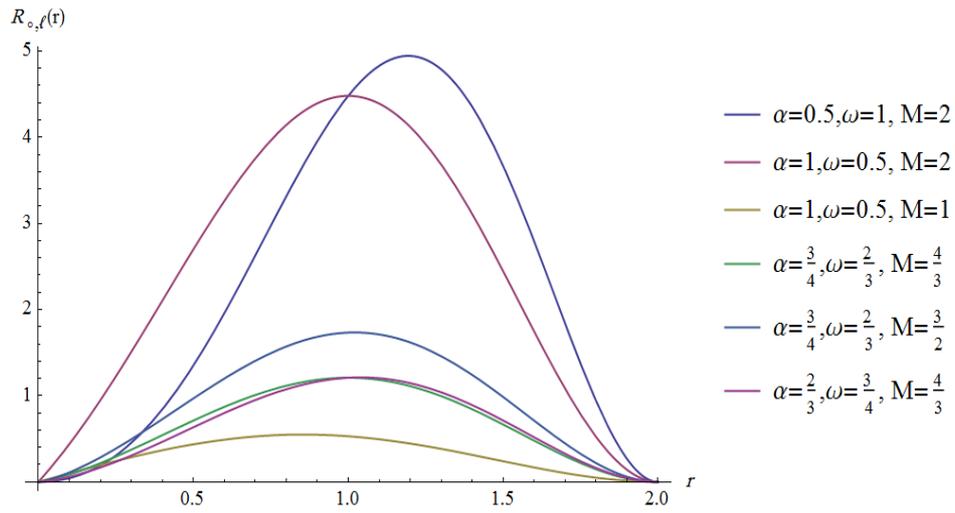

Fig. 2. The eigenfunctions for $b_1 = M, b_2 = \dfrac{q}{2}; b_3 = \dfrac{2m+\alpha}{2\alpha}; b_4 = \dfrac{3}{2}$

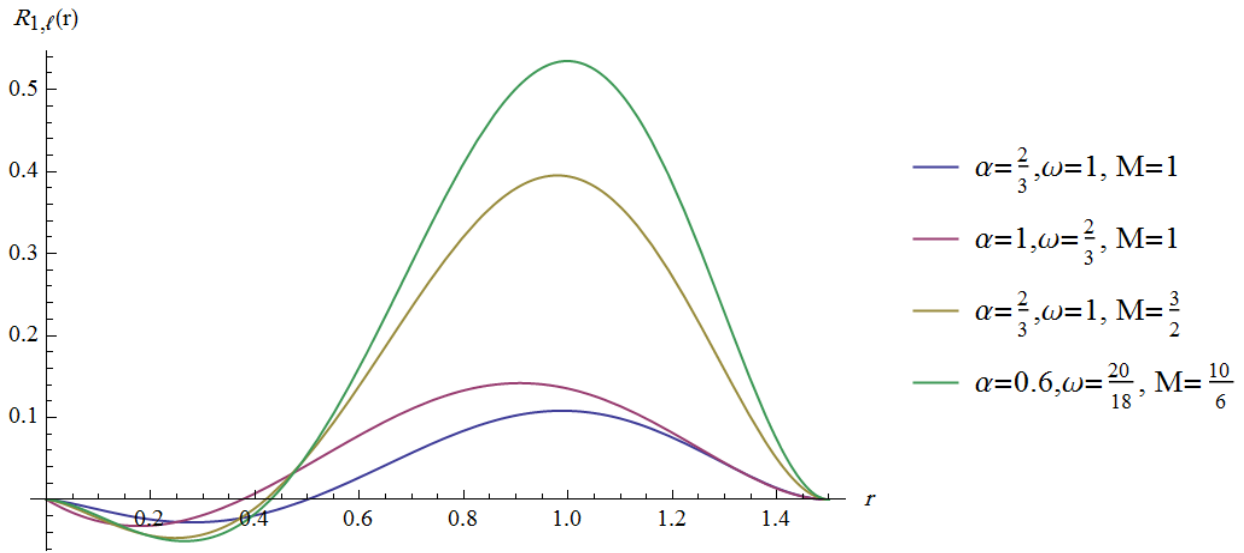

Fig. 3. The eigenfunctions of one-node states for $b_2 = \dfrac{q}{2}, b_1 = M, b_3 = \dfrac{2m+\alpha}{2\alpha}, b_4 = \dfrac{3}{2}$

Table 1. Energy of node-less states $b_1 = -M, b_2 = \frac{-q}{2}; b_3 = \frac{2m+\alpha}{2\alpha}; b_4 = \frac{3}{2}$ for different $\omega\alpha$

| $\omega\alpha$ | 0.001 | 0.002 | 0.003 | 0.004 | 0.005 |
|---|---|---|---|---|---|
| $E_{0,\ell}$ | -5.1090 <br> 5.0890 | -4.010 <br> 3.99 | -3.5690 <br> 3.5490 | -3.3266 <br> 3.3066 | -3.1722 <br> 3.1522 |
| $\omega\alpha$ | 0.006 | 0.007 | 0.008 | 0.009 | 0.01 |
| $E_{0,\ell}$ | -3.0650 <br> 3.0450 | -2.9861 <br> 2.9861 | -2.9254 <br> 2.9054 | -2.8774 <br> 2.8574 | -2.8384 <br> 2.8384 |

Table 2. Energy of node-less states $b_1 = -M, b_2 = \frac{-q}{2}; b_3 = \frac{2m+\alpha}{2\alpha}; b_4 = \frac{-1}{2}$ for different $\omega\alpha$

| $\omega\alpha$ | 0.001 | 0.002 | 0.003 | 0.004 | 0.005 |
|---|---|---|---|---|---|
| $E_{0,\ell}$ | -4.7004 <br> 4.6804 | -3.4741 <br> 3.4541 | -2.9539 <br> 2.9339 | -2.6557 <br> 2.6357 | -2.4594 <br> 2.4394 |
| $\omega\alpha$ | 0.006 | 0.007 | 0.008 | 0.009 | 0.01 |
| $E_{0,\ell}$ | -2.3194 <br> 2.2994 | -2.2138 <br> 2.1938 | -2.1313 <br> 2.1113 | -2.0448 <br> 2.0448 | -2.010 <br> 1.990 |

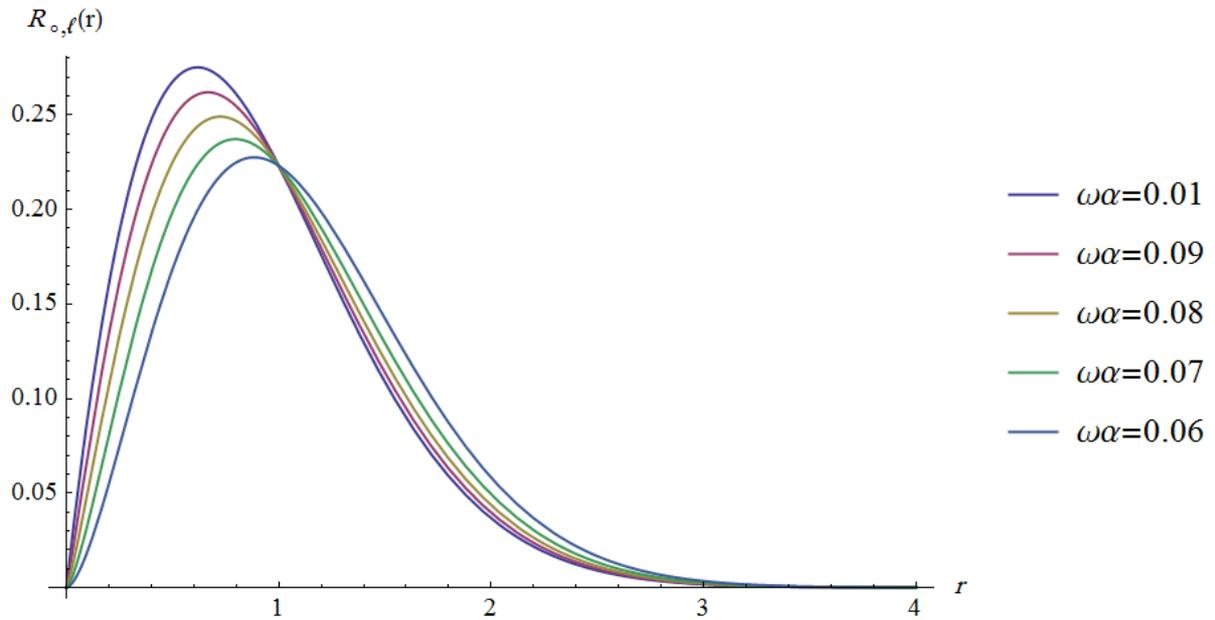

Fig. 4. The eigenfunctions for $b_1 = -M, b_2 = \frac{-q}{2}; b_3 = \frac{2m+\alpha}{2\alpha}; b_4 = \frac{3}{2}$ for different $\omega\alpha$

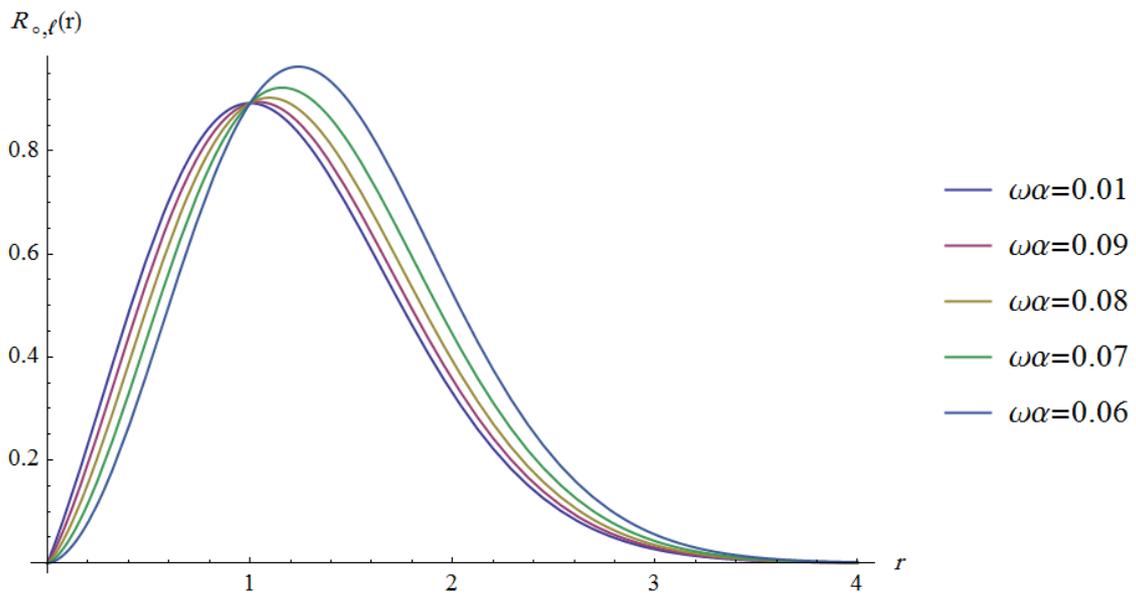

Fig. 5. The eigenfunctions for $b_1 = -M, b_2 = \dfrac{-q}{2}; b_3 = \dfrac{2m+\alpha}{2\alpha}; b_4 = \dfrac{-1}{2}$ for different $\omega\alpha$

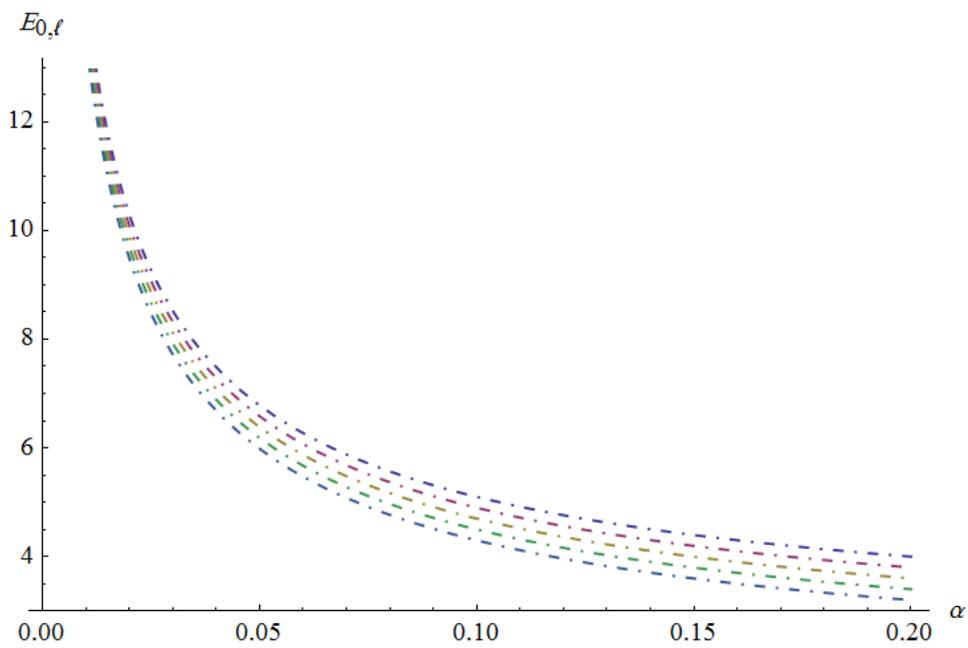

Fig. 6. The positive solution of energy as a function of $\alpha$ for $0 \leq \omega\alpha \leq 0.1$

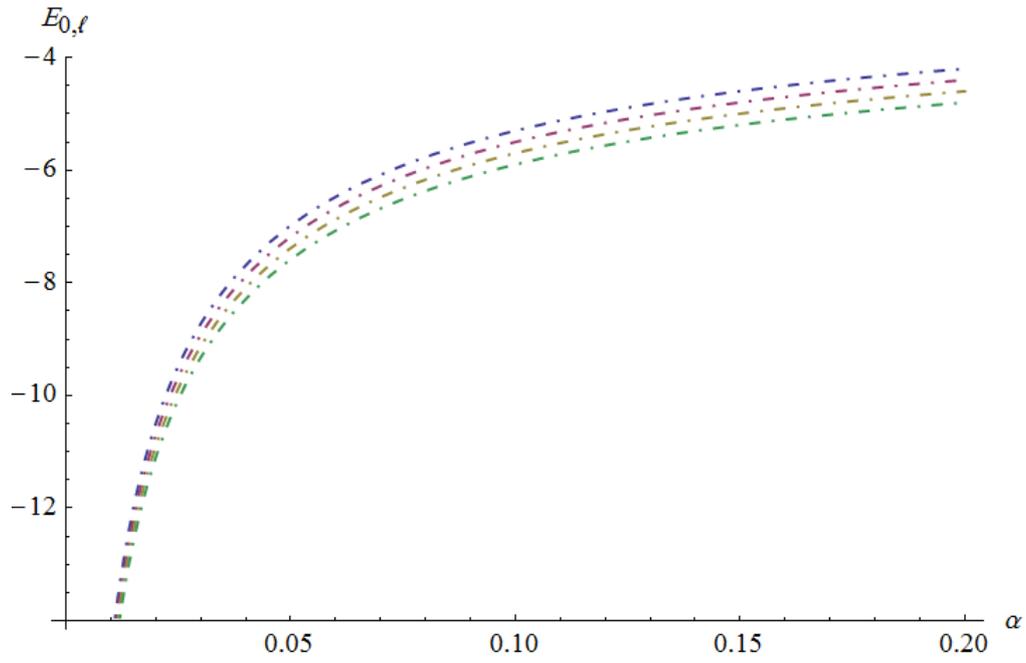

Fig. 7. The negative solution of energy as a function of $\alpha$ for $0 \leq \omega\alpha \leq 0.1$.

Table 3. Energy of one-node states for different $\omega\alpha$

| $\omega\alpha$ | 0.001 | 0.002 | 0.003 | 0.004 | 0.005 |
|---|---|---|---|---|---|
| $\alpha_1^1$ | 10.9091 | 5.8333 | 4.1025 | 3.2142 | 2.6666 |
| $E_{1,1}$ | -5.3015  5.2815 | -4.2526  4.2326 | -3.8397  3.8197 | -3.6155  3.5955 | -3.4742  3.4541 |
| $\omega\alpha$ | 0.006 | 0.007 | 0.008 | 0.009 | 0.01 |
| $\alpha_1^1$ | 2.2916 | 2.0168 | 1.8055 | 1.6374 | 1.5 |
| $E_{1,1}$ | -3.3765  3.3565 | -3.3050  3.2850 | -3.25039  3.23039 | -3.2072  3.1872 | -3.1722  3.1522 |

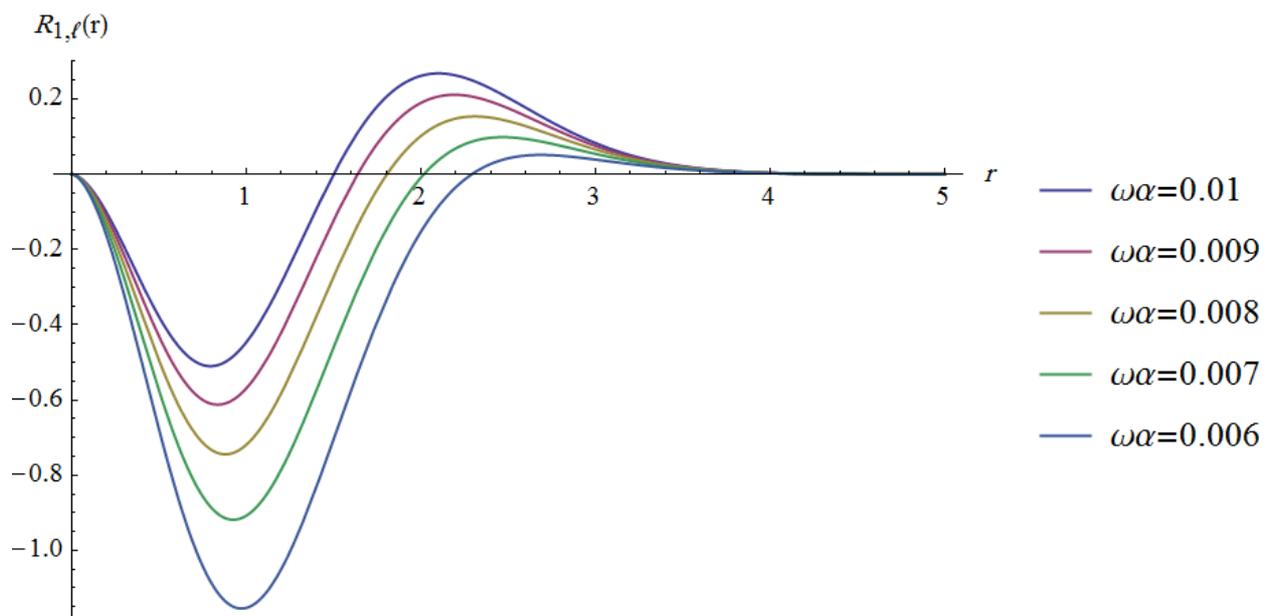

Fig. 8. The eigenfunctions of one-node states for different $\omega\alpha$

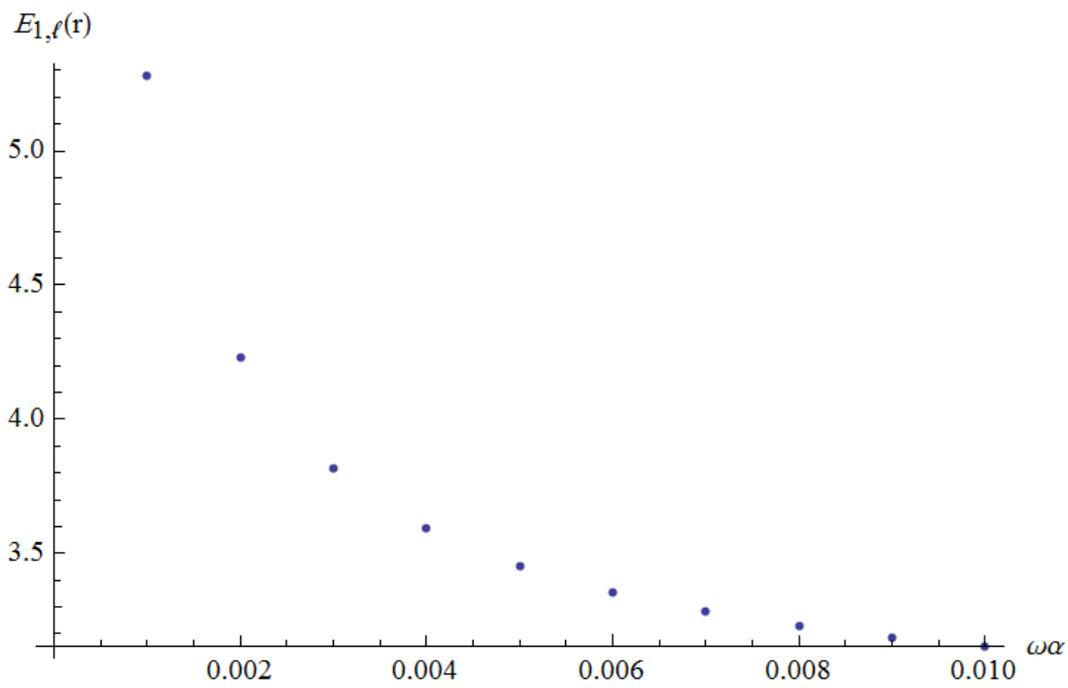

Fig. 9. The positive solution of energy as a function of $\omega\alpha$ for one-node states

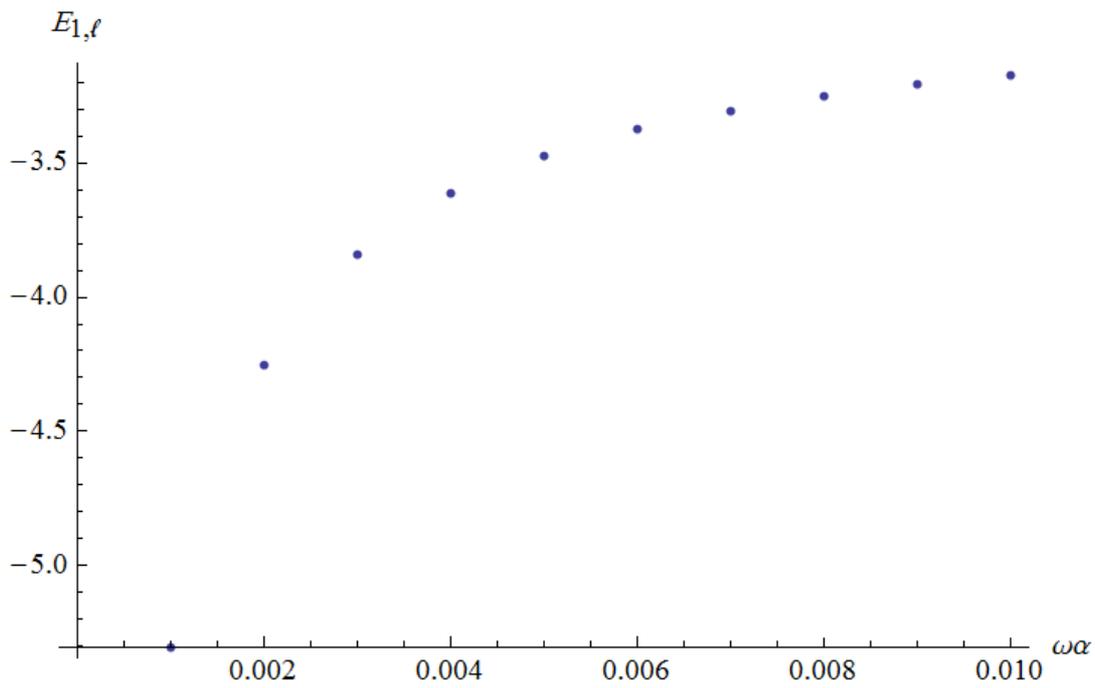

Fig. 10. The negative solution of energy as a function of $\omega\alpha$ for one-node states